\documentclass[aps,prl,twocolumn,superscriptaddress,nofootinbib,floatfix]{revtex4-2}
\usepackage[utf8]{inputenc}
\usepackage[T1]{fontenc}
\usepackage{amsmath,amssymb}
\usepackage{graphicx}
\usepackage{booktabs}

\newcommand{\RR}{{}^{*}\!R\,R}

\newcommand{\cpd}{\mathrm{cpd/kg/keV}}

\begin{document}

\title{Node-locked phase of annual modulations from the gravitational
chiral anomaly in the solar Kerr field}

\author{M.~Misiaszek}
\affiliation{Jagiellonian University, Faculty of Physics, Astronomy
and Applied Computer Science, M.~Smoluchowski Institute of Physics,
{\L}ojasiewicza 11, 30-348 Krak\'ow, Poland}

\date{\today}

\begin{abstract}
The leading parity-odd mass--spin curvature invariant of the solar
exterior, $P_{\odot}\equiv\RR\simeq
288\,G^{2}M_{\odot}^{2}a_{\odot}\cos\theta/c^{4}r^{7}$, changes sign
when the Earth crosses the solar equatorial plane and acts as the
geometric source of the gravitational chiral anomaly. We study two
minimal phenomenological responses to this structure: a long-memory
reservoir $Q\propto\int(P_{\odot}-\langle P_{\odot}\rangle)\,dt$ and
the local worldline derivative $D=u^{\mu}\nabla_{\mu}P_{\odot}$.
Both have an annual Fourier
phase fixed by ephemerides, $t^{*}=158.7$~d (June 7--8) or the
opposite branch $t^{*}=341.4$~d, with a calculable secular drift of
$+0.014$~d\,yr$^{-1}$ and an energy-independent geometric input
phase, while predicting different semiannual fractions, $3.75\%$ and
$15\%$, respectively. The response amplitudes and their microscopic
origin are not predicted; the phase is. Single-amplitude fits to the
digitized DAMA/LIBRA--phase2 1--3~keV residuals give
$\chi^{2}/\mathrm{dof}=62.7/51$ for the reservoir and $63.6/51$ for
the derivative, against $60.9/51$ for the standard-halo cosine. The
most precise published phase, $t^{*}=153.5\pm3.8$~d, lies
$0.3\sigma$ from the halo value and $1.4\sigma$ from the node-locked
one; phase metrology at the two-day level ($\simeq3\sigma$),
together with the phase-locked semiannual component, discriminates
between the two clocks.
\end{abstract}

\maketitle

\emph{Introduction.}---Annual-modulation searches rest on a
time-domain signature: a rate component
$S_{m}\cos\omega(t-t_{0})$ with a one-year period and a predicted
phase. In the standard halo model (SHM) the expectation
$t_{0}\simeq152.5$~d (June~2) follows from the solar motion toward
Cygnus combined with the Earth's orbital velocity
\cite{Drukier1986,Freese1988}. The DAMA/NaI and DAMA/LIBRA experiments
report a modulation of single-hit events in the low-energy region of
NaI(Tl) detectors satisfying this signature, most recently at
$15.3\sigma$ over a cumulative exposure of 3.40~t\,yr
\cite{DAMA2018,DAMA2026}, while the same-target experiments ANAIS-112
and COSINE-100 exclude a modulation of the DAMA amplitude at
$4.7\sigma$ (1--6~keV) \cite{CarlinPRL2025}; for a review of the
direct-detection landscape see Ref.~\cite{MisiaszekRossi}.

The June phase has long served as the principal argument for a
galactic origin of the effect: it matches no natural timescale of the
Solar System---perihelion falls on January~3, aphelion on July~4, and
seasonal backgrounds peak in local summer---so a signature peaking in
early June appeared explicable only by the Earth's velocity relative
to an external, galactic frame \cite{DAMA2018}. This argument by
elimination, however, overlooks a particularly simple Solar-System
structure with the appropriate calendar phase: the solar equatorial
plane. Newtonian gravity is blind to it---the
monopole field of a sphere retains no memory of the rotation
axis---but its signed north--south orientation is encoded naturally
in the parity-odd mass--spin sector of the solar curvature, through
the Chern--Pontryagin density
$P_{\odot}\propto\cos\theta$. The Earth crosses this plane on June~6--7
(ascending) and December~7 (descending); we show below that the fitted
phase of the resulting annual-modulation template lies, for one sign
of the coupling, six days from the dark-matter expectation.

The two hypotheses differ in the rigidity of their phase. The halo
value $t_{0}=152.5$~d presupposes a fully virialized, nonrotating
local halo; admitting bulk rotation, streams such as S1
\cite{OHare2018}, debris flows, or gravitational focusing by the Sun,
the predicted phase ranges over the entire year and acquires an energy
dependence \cite{Freese2005,Focusing}. The geometric phase derived in
this Letter is, by contrast, a two-element set fixed by the anomaly
equation and the ephemerides,
$t^{*}\in\{158.7~\mathrm{d},\,341.4~\mathrm{d}\}$; the only freedom is
the value of the coupling constant, and the phase is common to
observables coupled to an accumulated response
and to the worldline derivative of the anomaly
source. The construction uses nothing beyond the
rotation of the Sun and the geometry of the Earth's orbit, and it
predicts, in addition to the phase, its secular drift, its harmonic
content, and a set of null tests. Confronted with the DAMA data the
template is statistically indistinguishable from the SHM cosine, while
being falsifiable at the level of ephemeris precision.

At the level of a model-independent annual-modulation fit, both
descriptions contain an unknown amplitude. In the halo case it is set
by the particle coupling, mass, local density, and detector response;
in the present construction it is absorbed into the response
coefficient $\kappa$. The comparison made below is therefore a phase
comparison: the SHM clock is set by Galactic kinematics, whereas the
node-locked clock is set by the solar-rotation axis and ephemerides.

\emph{The phase, derived.}---Three ingredients lead from the solar
mass--spin field to a calendar date. First, for a massless Dirac
field on a classical gravitational background the axial current obeys
\cite{Kimura1969,DelbourgoSalam1972,EguchiFreund1976,AGW1984}
\begin{equation}
\nabla_{\mu}j^{\mu}_{5}
=\frac{1}{384\pi^{2}}\,\epsilon^{\mu\nu\rho\sigma}
R_{\mu\nu}{}^{\lambda\kappa}R_{\rho\sigma\lambda\kappa}
\equiv\frac{P}{192\pi^{2}} ,
\label{eq:anomaly}
\end{equation}
where
$P\equiv{}^{*}\!R_{\mu\nu\rho\sigma}R^{\mu\nu\rho\sigma}
=\tfrac12\,\epsilon^{\mu\nu\alpha\beta}
R_{\alpha\beta}{}^{\rho\sigma}R_{\mu\nu\rho\sigma}$;
a single chiral Weyl current carries one half of
the Dirac axial coefficient, with the sign fixed by its chirality,
and the orientation convention for $\epsilon^{\mu\nu\rho\sigma}$
fixes the overall signs of both $P$ and the anomaly. A nonvanishing
Pontryagin density acts as a local source term in the anomalous
divergence of the axial current, and $P$ is the unique independent
parity-odd scalar quadratic in the curvature and containing no
curvature derivatives. Throughout, we distinguish the geometric
pseudoscalar $P$, the reservoir response $Q$, and the
worldline-derivative response $D=u^{\mu}\nabla_{\mu}P$, defined
below.
Related consequences include neutrino studies in Kerr geometry
\cite{Unruh1973}, asymmetric neutrino emission by rotating black holes
\cite{LeahyUnruh1979,Vilenkin1979}, anomalous transport
\cite{Landsteiner2011}, and the chiral imprint of gravitational
radiation from binary mergers \cite{delRio2020}.

Second, in the leading Kerr-like mass--spin approximation to the
solar exterior, with
$M\equiv GM_{\odot}/c^{2}=1477$~m and
$a\equiv J_{\odot}/M_{\odot}c\simeq322$~m, both quadratic invariants
follow from the Weyl scalar
$\psi_{2}=-M/(r-\mathrm{i}a\cos\theta)^{3}$ \cite{Cherubini2002},
\begin{equation}
K+\mathrm{i}\,P=\frac{48M^{2}}{(r-\mathrm{i}a\cos\theta)^{6}}
\;\Rightarrow\;
P_{\odot}\xrightarrow[r\gg a]{}\frac{288M^{2}a\cos\theta}{r^{7}}
\label{eq:kerr}
\end{equation}
(the overall sign of $P$ is an orientation convention; its
equatorial antisymmetry is not). The density is exactly odd about the solar equator and integrates to
zero over any sphere; a localized detector, however, samples a
definite sign.

Third, the Earth's polar angle with respect to the solar rotation
axis (inclined $i=7.25^{\circ}$ to the ecliptic) oscillates with the
heliographic latitude, $\cos\theta_{\oplus}=\sin B(t)$, so the sampled
density is proportional to $\sin B(t)/r(t)^{7}$: it vanishes and
changes sign at the crossings of the solar equatorial plane
($\sim$June~7 ascending, $\sim$December~7 descending). The local
anomaly source has one sign throughout June--December and the
opposite sign during the other half of the year. In the long-memory reservoir limit one may
phenomenologically write
$Q(t)\propto\int^{t}[P_{\odot}(t')-\langle P_{\odot}\rangle]\,dt'$;
this response is extremal close to the crossings of the solar
equatorial plane.

On the elliptic orbit the exact waveform is not a pure cosine; a
cosine fit, however---the operation performed by every modulation
analysis---measures the phase of the annual Fourier component, and
the leading $O(e)$ eccentricity distortion is semiannual and hence
orthogonal to the annual component over a complete cycle.
Throughout this Letter ``phase'' denotes the phase of the annual
Fourier component, not the date of the maximum of the full
nonsinusoidal waveform: the exact maxima occur on day~156.6 for the
reservoir and day~147.0 for the derivative, while
both annual fundamentals share $t^{*}=158.66$~d on the June
branch. The fitted phase is therefore the
date at which the Earth's mean longitude crosses the node of the solar
equator: day 158.7 (June 7--8) or day 341.4 (December 7--8)---the
difference between the Carrington node longitude
and the longitude of perihelion. For the reservoir response the June
branch corresponds to a rate-suppressing coupling; the mapping
inverts for the derivative response. Numerically, a free-phase cosine fit
to the exact Kepler waveform returns 158.66~d, and 158.6~d when the
waveform is evaluated on the actual DAMA live-time grid (52 bins with
calibration gaps, published uncertainties as weights). No velocity
distribution, halo model, or astrophysical assumption enters this
date.

The six-day separation from the dark-matter expectation has a simple
origin: it is the angle between two unrelated directions on the sky.
The halo phase (June~2) is set by the direction of the solar motion
toward Cygnus---galactic kinematics; the node date (June~7--8) by the
orientation of the solar rotation axis---a relic of the protosolar
angular momentum. The two dates lie $6.2$~days---$6.1^{\circ}$ of mean
longitude---apart. This coincidence is why the node date remained
hidden inside the ``June window'' for three decades: the standard
argument by elimination checked perihelion, aphelion, and the
seasons, and an early-June phase appeared uniquely galactic. Had the
solar axis pointed elsewhere, the two hypotheses would have been
separable from the first annual cycle.

\emph{Predictions.}---The amplitude of the coupling of an accumulated
chiral background to weak-interaction--driven rates is not predicted
here: within perturbative curved-space quantum field theory it is
unobservably small for point-vertex processes, and for
extended-fermion scenarios no reliable estimate exists. What is
predicted, with zero adjustable parameters, is the waveform. Writing
the rate as $R(t)=R_{0}[1+\kappa f(t)]$, the periodic reservoir
template is the time integral of the mean-subtracted sampled density
along the exact Earth orbit,
\begin{equation}
s(t)=\frac{\sin B(t)}{r(t)^{7}}
-\Big\langle\frac{\sin B}{r^{7}}\Big\rangle_{\rm yr},
\qquad
f(t)=\mathcal{N}\int^{t}\!s(t')\,dt',
\label{eq:template}
\end{equation}
with $\mathcal{N}$ fixing the arbitrary normalization---computable
from the ephemeris alone. The subtraction removes the small constant
(time-averaged) component generated by the eccentric $r^{-7}$
weighting and has no
effect on the annual Fourier phase; physically, a finite relaxation
time converts this constant source term into a steady offset absorbed
in $R_{0}$.
Its properties: (i) cosine-fit phase equal to the node-locked date,
the branch selected by the sign of $\kappa$; (ii) secular phase drift
$+0.0142$~d\,yr$^{-1}$, the sidereal--tropical year difference, since
the Carrington node motion is pure equinox precession; (iii) an
energy-independent phase of the geometric input, this being a
property of the external geometry; an observed energy dependence
would indicate an energy-dependent material response or a
superposition of modulated components; (iv) a
semiannual harmonic of $3.75\%$ of the fundamental, phase-locked,
generated by orbital eccentricity through the $r^{-7}$ weighting and
the equation of the center, with an amplitude that grows with the
radial exponent and therefore provides an internal consistency
test; and (v) a common phase in both terrestrial hemispheres,
$B(t)$ being a property of the Earth's position. A finite chirality relaxation time
$\tau_{5}$ interpolates the phase between the source limit
($t^{*}\simeq251$~d) and the reservoir limit (the node-locked date)
through $\tan\Delta\phi=\omega\tau_{5}$, where $\Delta\phi$ is the
lag of the response behind the source waveform; the node-locked value
applies to the reservoir and derivative limits studied here, and a
generic causal response kernel need not retain it exactly.

The reservoir is not the only simple linear functional that inherits
the node-locked annual phase. A second limiting response is the
worldline derivative of the anomaly source,
\begin{equation}
D(t)\equiv u^{\mu}\nabla_{\mu}P_{\odot}
=\frac{dP_{\odot}}{d\tau}\propto
\frac{d}{dt}\!\left[\frac{\sin B(t)}{r(t)^{7}}\right].
\label{eq:derivative}
\end{equation}
It measures the rate at which the detector traverses the parity-odd
solar curvature structure rather than an accumulated axial reservoir.
Moreover, for a stationary, axisymmetric terrestrial self-field the
laboratory four-velocity is built from Killing vectors, so
$u^{\mu}\nabla_{\mu}P_{\oplus}\simeq0$: the dominant constant Earth
contribution drops out of $D$, leaving residual terrestrial gradients
to diurnal and sidereal frequencies rather than to the annual
node-locked template.
For the annual fundamental, integration and differentiation generate
opposite quadratures of the same source, so both responses have the
same annual Fourier phase,
$t^{*}\in\{158.7,\,341.4\}$~d, the June branch corresponding to
$\kappa<0$ for $Q$ and $\kappa_{D}>0$ for $D$. Their full waveforms
differ: integration suppresses the $n$th harmonic by $1/n$, whereas
differentiation enhances it by $n$, giving
$(A_{2}/A_{1})_{Q}=3.75\%$ and $(A_{2}/A_{1})_{D}\simeq15\%$ and the
waveform maxima quoted above.

\emph{Confrontation with DAMA/LIBRA.}---The 52 single-hit residuals of
DAMA/LIBRA--phase2 in the 1--3~keV interval (1.13~t\,yr) were
digitized from the vector PostScript of Ref.~\cite{DAMA2018}. The
extraction reproduces the published analysis: null hypothesis
$\chi^{2}/\mathrm{dof}=126.6/52$ (published $127.3/52$); SHM cosine
with $t_{0}=152.5$~d, $A=0.0179\pm0.0022~\cpd$ and $\chi^{2}=60.9/51$
(published $0.0184\pm0.0023$, $61.3/51$). The comparison is made in
the same one-amplitude sense in which DAMA quotes the
model-independent annual modulation: the period and phase define the
clock, while the amplitude is fitted from the residuals. Thus the SHM
cosine and the two node-locked responses to the solar Kerr
density---the reservoir $Q\propto\int(P_{\odot}-\langle
P_{\odot}\rangle)\,dt$ and the worldline derivative
$D=u^{\mu}\nabla_{\mu}P_{\odot}$---enter the analysis
on equal footing, differing only in the predicted waveform (common
phase $t^{*}=158.7$~d, opposite June signs $\kappa<0$ and
$\kappa_{D}>0$, and semiannual fractions $3.75\%$ and $15\%$).
Both response templates are normalized to unit amplitude of their
annual Fourier component, so the fitted coefficients in
Table~\ref{tab:fits} are directly comparable annual rate amplitudes;
the predicted distinction resides in the higher harmonics.
Table~\ref{tab:fits} lists
single-amplitude fits, with per-cycle constant subtraction applied to
the templates exactly as applied by DAMA to the data. These fits do
not establish a microscopic gravitational coupling; they test whether
the temporal shapes implied by the two minimal response operators are
compatible with the published residuals after one overall rate
coefficient is fitted.

\begin{table}[b]
\caption{\label{tab:fits}Single-amplitude fits to the digitized
DAMA/LIBRA--phase2 (1--3~keV) residuals. Templates are normalized to
unit amplitude of their annual Fourier component.}
\begin{ruledtabular}
\begin{tabular}{lcc}
template & coeff.\ ($\cpd$) & $\chi^{2}$/dof\\
\colrule
SHM cosine ($t_{0}{=}152.5$~d) & $0.0179\pm0.0022$ & $60.9/51$\\
source limit ($\simeq$251~d) & $-0.0036\pm0.0023$ & $124.1/51$\\
reservoir, $\kappa{<}0$ (158.7~d) & $-0.0176\pm0.0022$ & $62.7/51$\\
derivative, $\kappa_{D}{>}0$ (158.7~d) & $0.0172\pm0.0022$ & $63.6/51$\\
\end{tabular}
\end{ruledtabular}
\end{table}

\begin{figure}[t]
\includegraphics[width=\columnwidth]{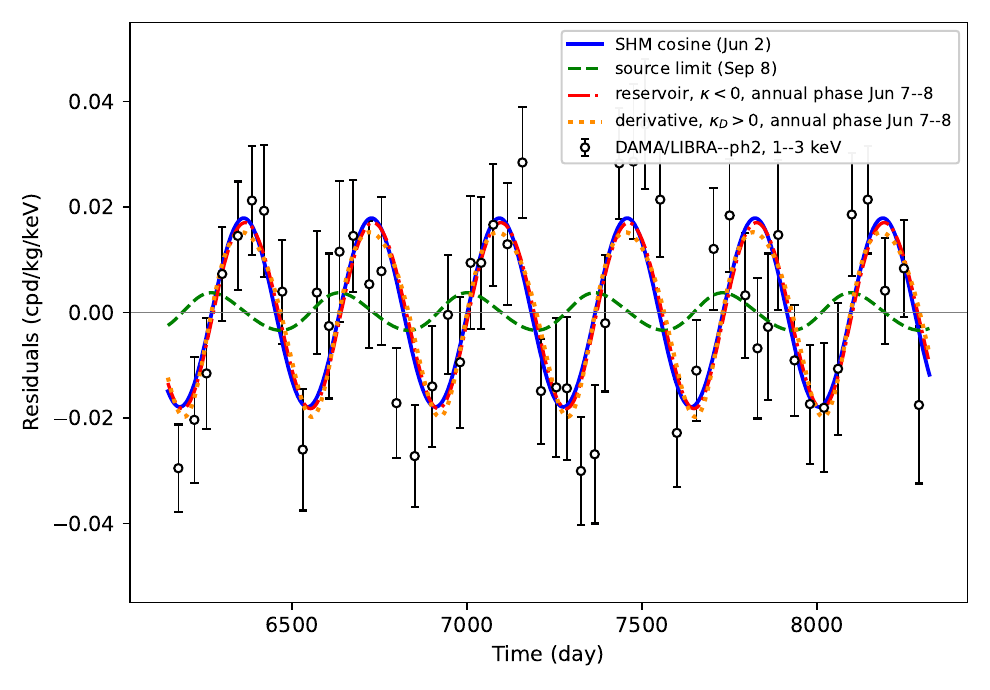}
\caption{\label{fig:fit}Digitized DAMA/LIBRA--phase2 (1--3~keV)
residuals with the SHM cosine, the source-limit template, and the two
node-locked templates: the reservoir with $\kappa<0$ and the worldline
derivative with $\kappa_{D}>0$. Template amplitudes are fitted with
the same per-cycle constant subtraction applied by DAMA to the
residuals; dates in the legend denote the phase of the annual Fourier
component.}
\end{figure}

The source-limit template is strongly disfavored as the sole
description of the observed modulation,
consistently with the null quadrature component
$Z_{m}=-0.0003\pm0.0008~\cpd$ (2--6~keV, full exposure) reported by
DAMA \cite{DAMA2018}, whose phase nearly coincides with that of the
source template. The two node-locked linear responses remain
competitive with SHM: the reservoir ($\kappa<0$, accumulated
chirality suppressing the rate) gives $\chi^{2}=62.7/51$
[$\Delta\chi^{2}(Q-\mathrm{SHM})=1.8$], and the derivative
($\kappa_{D}>0$, rate enhanced by the crossing speed) gives
$\chi^{2}=63.6/51$ [$\Delta\chi^{2}(D-\mathrm{SHM})=2.7$]
(Fig.~\ref{fig:fit}). At the present precision the data separate
neither $P_{\odot}$ template from the SHM cosine nor the two
$P_{\odot}$ responses from each other.

\begin{figure}[t]
\includegraphics[width=\columnwidth]{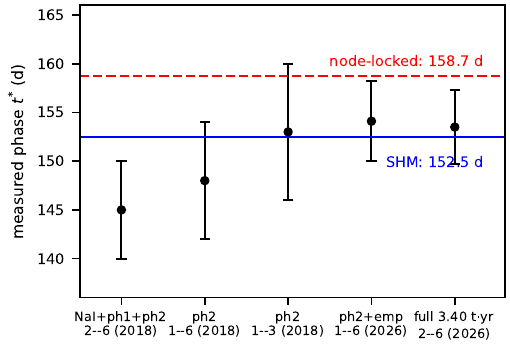}
\caption{\label{fig:phase}Published free-phase measurements of the
2018 and 2026 DAMA releases \cite{DAMA2018,DAMA2026} compared with the
SHM expectation (solid) and the node-locked prediction (dashed). At the present precision every
determination is compatible with both.}
\end{figure}

The published free-phase determinations of the two primary data
releases are summarized in Fig.~\ref{fig:phase}. The 2018 release
gives $t^{*}=145\pm5$~d for the full 2.46~t\,yr exposure (2--6~keV)
and $148\pm6$, $153\pm7$~d for phase2 alone (1--6, 1--3~keV)
\cite{DAMA2018}; the 2026 release, including phase2-empowered, gives
$154.1\pm4.1$~d (2.07~t\,yr, 1--6~keV) and $153.5\pm3.8$~d for the
full 3.40~t\,yr exposure (2--6~keV) \cite{DAMA2026}. The most precise
value lies $0.3\sigma$ from the SHM phase and $1.4\sigma$ from
the node-locked date: at the present
$4$--$5$~d precision both hypotheses are compatible with all
determinations; a two-day determination would separate them at the
$3\sigma$ level.

Two consistency conditions constrain the interpretation exactly as
they constrain dark matter. First, DAMA observes no modulation of the
bulk rate in the phase2-empowered data ($R_{90}$ amplitude
$-0.011\pm0.073$~cpd/kg; 6--10~keV amplitude
$0.0011\pm0.0011~\cpd$ \cite{DAMA2026}), and decay-constant
metrology bounds generic annual variations of radioactivity at the
$10^{-4}$--$10^{-5}$ level \cite{Pomme2016}; the coupling must
therefore be selective to the process populating the 0.5--6~keV
single-hit window---not excluded a priori: weak decays are parity
violating, and a periodic redistribution of a low-energy $\beta$
spectrum (e.g.\ cosmogenic $^{3}$H) could shift counts across the
window without modulating the total decay constant. Second, the solar template has
the same geometric phase at Gran Sasso, Canfranc, and Yangyang, so
ANAIS-112 and COSINE-100 \cite{CarlinPRL2025} test the phase
directly; comparing amplitudes, however, requires the same carrier
and response in setups whose low-energy backgrounds (including
$^{3}$H) differ.

\emph{Discussion.}---The SHM also predicts a date, but that date is
inherited from an assumed velocity distribution and is known to shift
with halo substructure and with gravitational focusing, acquiring in
general an energy dependence \cite{Freese2005,Focusing}. The phase of
the geometric input, by contrast, is fixed by
ephemerides---computable to $0.1$~d over the next century---and is
invariant under the choice of linear response (reservoir or worldline
derivative); additional phase freedom can arise only from a
nontrivial material response time or a superposition of response
channels. The two
hypotheses are separated by $6.2$~d in phase, by the presence or
absence of an energy dependence of $t^{*}$, and by a phase-locked
semiannual harmonic---$3.75\%$ for the reservoir and $15\%$ for the
derivative response. A two-day phase determination---within reach of
a per-era reanalysis of the DAMA archive (currently $3.8$~d) and of
the forthcoming SABRE North--South pair
\cite{SABRENorth2022,SABRESouthTDR}---would discriminate between the
standard-halo June-2 clock and the node-locked June~7--8 clock,
while testing the energy independence of the geometric input phase,
the hemisphere-common phase, and the fixed semiannual harmonic.

\begin{acknowledgments}
The work was supported by the National Science Centre,
Poland [Grant No.\ 2025/59/B/ST2/02359].
\end{acknowledgments}

\end{document}